\documentclass[final,12pt]{elsarticle}
\usepackage{amsthm}
\usepackage{amsmath}
\usepackage{upref}
\usepackage{amsfonts}
\usepackage{amssymb}
\usepackage{mathpazo}
\usepackage{times}

\newcommand{\N}{\mathbb{N}}

\newcommand{\mathset}[1]{\left\{#1\right\}}
\newcommand{\abs}[1]{\left|#1\right|}
\newcommand{\ceilenv}[1]{\left\lceil #1 \right\rceil}
\newcommand{\floorenv}[1]{\left\lfloor #1 \right\rfloor}
\newcommand{\parenv}[1]{\left( #1 \right)}
\newcommand{\sparenv}[1]{\left[ #1 \right]}

\newcommand{\oodd}{1_{\mathrm{odd}}}
\DeclareMathOperator{\per}{per}
\DeclareMathOperator{\ord}{ord}

\newtheorem{conjec}{Conjecture}

\newtheorem{theorem}{Theorem}
\newtheorem{lemma}[theorem]{Lemma}
\newtheorem{corollary}[theorem]{Corollary}
\newdefinition{definition}{Definition}
\newdefinition{remark}{Remark}
\newdefinition{const}{Construction}
\newdefinition{example}{Example}

\begin{document}

\journal{Journal of Combinatorial Theory Ser.~A}

\begin{frontmatter}

\title{On Optimal Anticodes over
Permutations with the Infinity Norm}

\author{Itzhak Tamo}
\ead{tamo@ee.bgu.ac.il}
\address{Department of Electrical and Computer Engineering, Ben-Gurion
University, Israel} 
\author{Moshe Schwartz}
\ead{schwartz@ee.bgu.ac.il}
\address{Department of Electrical and Computer Engineering, Ben-Gurion
University, Israel} 

\address{}

\begin{abstract}
Motivated by the set-antiset method for codes over permutations under
the infinity norm, we study anticodes under this metric. For half of
the parameter range we classify all the optimal anticodes, which is
equivalent to finding the maximum permanent of certain
$(0,1)$-matrices. For the rest of the cases we show constraints on the
structure of optimal anticodes.
\end{abstract}


\end{frontmatter}

\section{Introduction}

Perhaps the first to pioneer the study of permutation theory from a
coding perspective were Chadwick and Kurz \cite{ChaKur69}, and Deza
and Frankl \cite{DezFra77}. The object under study in their work,
called a \emph{code} or a \emph{permutation array}, is a set of
permutations for which any two distinct members are at distance at
least $d$ apart. Codes over permutations have attracted recent
interest due to power line communication
\cite{VinHaeWad00,ChuColDuk04,FuKlo04,LinTsaTze08} and storage schemes
for flash memories
\cite{JiaSchBru10,TamSch10,KloLinTsaTze10,BarMaz10}. Moreover, in
\cite{Bai09,TamSch10} permutation groups were considered as
error-correcting codes.
 
Since a distance measure is involved in the definition of a code, a
proper choice of metric is important. There are many well-known
metrics over the symmetric group $S_n$ (see \cite{DezHua98}), of which
the Hamming metric is by far the most studied. However, the infinity metric
induced by the infinity norm has received recent interest due to its
applications \cite{TamSch10,KloLinTsaTze10}, and will serve as the
metric of choice in this work. The $\ell_\infty$-distance
between two permutations $f,g\in S_n$ is defined as
\[d(f,g)=\max_{1\leq i \leq n}\abs{f(i)-g(i)}.\]

In analogy to the definition of a code, a subset $A\subseteq S_n$ is
an \emph{anticode} with maximal distance $d$, if any two of its
members are at distance at most $d$ apart.  In the context of
coding theory, the first use of anticodes was in \cite{SolSti65,Far70}
(see also \cite{MacSlo78} for an overview).  The anticodes were in
fact multisets (allowing repeated words), and were used to construct
codes that attain the Griesmer bound with equality. As a purely
combinatorial question, anticodes (though not under this name) appear
in earlier works, such as the celebrated Erd{\H{o}}s-Ko-Rado Theorem
on $t$-intersecting families \cite{ErdKoRad61}. Many other variations
on the ambient space and distance measure have created a wealth
of anticodes, see for example
\cite{FraWil86,MarZhu95,AhlKha98,SchEtz02,EtzSchVar05}.

The following theorem, which is sometimes referred to as the
\emph{set-antiset theorem}, motivates us to explore anticodes of
maximum size. This theorem was also used before over different
spaces and distance measures (see \cite{Del73,DezFra77,AhlAydKha01,TamSch10}).

\begin{theorem}[{\cite[Theorem 13]{TamSch10}}]
\label{set-antiset} 
Let $C,A\subseteq S_n$ be a code and an anticode under the
$\ell_\infty$-metric, with minimal distance $d$
and maximal distance $d-1$, respectively. Then
\[\abs{C}\cdot|A|\leq\abs{S_n}=n!.\]
\end{theorem}
 
It should be noted that balls are just a special case of anticodes,
since a ball of radius $r$ is an anticode with maximal distance $2r$.
The size of balls in $S_n$ under the $\ell_\infty$-metric
has been studied in \cite{Sch09,Klo09}.

It is well known (see \cite{DezHua98})
that the $\ell_\infty$-metric over $S_n$ is right
invariant, i.e., for any $f,g,h\in S_n$, $d(f,g)=d(fh,gh)$.  Hence,
w.l.o.g., one can assume that any code or anticode contains the
identity permutation simply by taking a translation, and we shall
assume so throughout the paper.

Any anticode $A\subseteq S_n$ of maximum distance $d-1$ defines a
$(0,1)$-matrix $A^*=(a_{i,j})$ of order $n$, for which $a_{i,j}=1$ iff
there exists $f\in A$ such that $f(i)=j$. We note that $A^*$ has the
property that if $a_{i,j}=1$ then $\abs{i-j}\leq d-1$. Moreover, the
$(0,1)$-matrix $A^*=(a_{i,j})$ defines an anticode $B$ with maximum
distance $d-1$ by
$B=\mathset{f \in S_n  :  a_{i,f(i)}=1\ \text{ for all } i\in [n]}$.
Note that $A\subseteq B$ and that the size of $B$ is the
permanent of the matrix $A^*$, which is defined
by
\[\per(A^*)=\sum_{f \in S_n}\prod_{i=1}^n a_{i,f(i)}=\abs{B}.\]

Let $\Gamma_n^d$ denote the set of $(0,1)$-matrices of order $n$ with
exactly $d$ non-zero entries in each row which form a contiguous
block. Let $A^*$ be a $(0,1)$-matrix defined by an anticode $A$, then
by the previous observation, the set of non-zero entries in $A^*$ is a
subset of the non-zero entries of some matrix in $\Gamma_n^d$. Thus,
every anticode $A$ with maximum permanent is equivalent to a matrix
$A^*\in \Gamma_n^d$.

The goal of this paper is to study the structure of matrices that
attain the maximum permanent, i.e., the set of matrices
\[M_n^d=\mathset{A\in \Gamma_n^d  :  \per(A)\geq \per(B) \text{ for all } B\in
\Gamma_n^d},\]
and to calculate the value of the maximum permanent.

Similar questions regarding the value of the maximum permanent and the
matrices that attain it, have been studied for other sets of matrices. Perhaps
the most related is the study of constant line-sum $(0,1)$-matrices, in which
the number of non-zero entries in each row and each column is equal. This
is still an open problem, first stated by Minc \cite{Min78}, and more
recently studied by Wanless \cite{Wan99}.

The problem was partly solved, both for constant line-sum $(0,1)$-matrices,
and for the matrices studied in this paper, $\Gamma_n^d$, by Br{\'e}gman
\cite{Bre73}, who showed that if $A$ is a $(0,1)$-matrix of order $n$
and row sums $d_1,d_2,\dots,d_n$ then
\begin{equation}
\label{bregman inequality}
\per(A)\leq \prod_{i=1}^n(d_i!)^\frac{1}{d_i}.
\end{equation} 
Moreover, equality holds iff $d_1=d_2=\dots=d_n=d$ and $A$ is a direct
sum of $1_{d\times d}$ matrices, where $1_{d\times d}$ is the all-ones
square matrix of order $d$. Our results focus
on the case where $d$ does not divide $n$.  The main results of this
paper are:
\begin{itemize}
\item 
  For $d > \frac{n}{2}$ let $A\in M_n^d$, then up to row and
  column permutations, in the first $\floorenv{\frac{n}{2}}$
  rows, the non-zero blocks are flushed to the left, and in the last
  $\floorenv{\frac{n}{2}}$ rows, they are flushed to the right.
  For $n$ even this looks like
\newcommand{\pz}{\phantom{0}} \newcommand{\po}{\phantom{1}}
\[
A=\begin{pmatrix}
\begin{array}{|ccc|}
\hline
\po & \po & \po \\
\po & 1_{(n/2)\times d} & \po \\
\po & \po & \po \\
\hline
\end{array}
\begin{array}{|c|}
\hline
\pz \\
0_{(n/2)\times (n-d)} \\
\pz \\
\hline
\end{array} \\
\begin{array}{|c|}
\hline
\pz \\
0_{(n/2)\times (n-d)} \\
\pz \\
\hline
\end{array}
\begin{array}{|ccc|}
\hline
\po & \po & \po \\
\po & 1_{(n/2)\times d} & \po \\
\po & \po & \po \\
\hline
\end{array}

\end{pmatrix},
\]
where $1_{i\times j}$ (respectively, $0_{i\times j}$) denotes the all-ones
(respectively, all-zeros) matrix of size $i\times j$. When
$n$ is odd, note that the position of the non-zero block in
the middle row is unconstrained.
Thus, for any $A\in M_n^d$,
\[\per(A)=\binom{2d-n}{\floorenv{ \frac{2d-n}{2}}}
\parenv{\floorenv{\frac{n}{2}}}!
\parenv{\ceilenv{\frac{n}{2}}}!.\]
           
\item 
  For $d < \frac{n}{2}$ we give some results based on results of
  Wanless \cite{Wan99}, adjusted to our case.  We show that any
  $A\in M_n^d$ has a certain structure, and for sufficiently-large $n$, $A$
  satisfies some periodic property.
\end{itemize}

The rest of the paper is organized as follows. In Section
\ref{sec:full classification} we focus on the case of
$d > \frac{n}{2}$, classify precisely all the optimal
anticodes up to isomorphism, and calculate their size. We proceed in
Section \ref{sec:assymptotic results} to the case of $d < \frac{n}{2}$
and present some asymptotic
results regarding the structure of the optimal anticodes.  We conclude
in Section \ref{sec:conclusion} with a summary of the results and
short concluding remarks.

\section{Full Classification for $d > \frac{n}{2}$}
\label{sec:full classification}

We consider the case of
$n=d+r$,  where $0<r<d$. Let $A\in \Gamma_{d+r}^d$,
$A=(a_{i,j})_{i,j\in[n]}$, and for any $i\in[d+r]$, we define
$x_i=\min{\mathset{j:a_{i,j}=1}}$, i.e, the left-most column of the non-zero
block in row $i$. Since the permanent is preserved under column and
row permutations we can assume w.l.o.g.~that $x_i\leq x_j$
for all $i\leq j$.

It can be seen that $A$ is defined uniquely by the vector
$(x_1,x_2,\dots,x_{d+r})$, so by abuse of notation we will sometimes
write 
\[A=(x_1,x_2,\dots,x_{d+r})\]
and also write 
\[\per(A)=\per(x_1,x_2,\dots,x_{d+r})=\per (\{x_i\}).\]
Also, for any $i,j\in[d+r]$ we define $\per(A_{i,j})$ to be
the permanent of $A$ after deleting row $i$ and column $j$.
 
In addition, for each $i\in[d+r]$, we define
\[i^*=\max{\mathset{k:x_k=x_i}}, \qquad\qquad i_*=\min{\mathset{k:x_k=x_i}}.\]
For all $i\in[d+r]$ such that $x_i\leq r$ we define the following operator
\[\per_i^+(\{x_k\})=\per(x_1,\dots,x_{i^*-1},x_{i^*}+1,x_{i^*+1},\dots,x_{d+r}).\]
If $x_i<x_{i+1}\leq r$ we define
\[\per_i^{++}(\{x_k\})=\per_i^+(\per_i^+(\{x_k\})).\]
Finally, in the same manner, for all $i\in[d+r]$ such that $2\leq x_i$
\[\per_i^-(\{x_k\})=\per(x_1,\dots,x_{i_*-1},x_{i_*}-1,x_{i_*+1},\dots,x_{d+r}).\]
If $2\leq x_{i-1}< x_i$ we define
\[\per_i^{--}(\{x_k\})=\per_i^-(\per_i^-(\{x_k\})).\]

\begin{lemma}
\label{lemma:first lemma}
For any $1\leq m\leq n \leq r$ 
\[\mathset{i:a_{i,m}=1}\subseteq\mathset{i:a_{i,n}=1},\]
and for any $d+1\leq m\leq n \leq d+r$ 
\[\mathset{i:a_{i,n}=1}\subseteq \mathset{i:a_{i,m}=1}.\]

\end{lemma}

\begin{proof}
Let $1\leq m\leq n \leq r$, and let $i\in \mathset{i:a_{i,m}=1}$. Since the
length of the non-zero block is $d$, for any $m\leq k \leq d$ we have
$a_{i,k}=1$, and in particular, for $k=n$ we get $a_{i,n}=1$.
Therefore, $i\in\mathset{i:a_{i,n}=1}$, and this proves the first part of the
lemma. The second part of the lemma follows easily from the symmetry
of the problem, i.e., rotating $A$ and using the first part of the
proof.
\end{proof} 

\begin{corollary}
\label{cor:first cor}
If $1\leq m\leq n\leq r$, then for any $i\in[d+r]$ we
get 
\[\per(A_{i,m})\geq \per(A_{i,n}).\]
If $d+1\leq m\leq n\leq d+r$, then for any $i\in[d+r]$ we get 
\[\per(A_{i,m})\leq \per(A_{i,n}).\]
\end{corollary}

\begin{proof}
For the first claim of the corollary, by Lemma \ref{lemma:first lemma}
we have
$\mathset{k:a_{k,m}=1}\subseteq \mathset{k:a_{k,n}=1}$.
Therefore, $A_{i,n}$ has the same columns as $A_{i,m}$ except for
one column in $A_{i,m}$ that has a superset of the $1$'s of the
corresponding column in $A_{i,n}$. Then we conclude
that $\per(A_{i,m})\geq \per(A_{i,n})$. The second claim of the
corollary is again proved by rotating $A$ and applying the first part
of the proof.
\end{proof}

\begin{lemma}
\label{stam}
Let $i$ be such that $x_i\leq r$, then
\[\per_i^+(\{x_k\})=\per(\{x_k\})+\per(A_{i^*,x_{i^*}+d})-\per(A_{i^*,x_{i^*}}).\]
Let $i$ be such that $2\leq x_i$ then
\[\per_i^-(\{x_k\})=\per(\{x_k\})+\per(A_{i_*,x_{i_*}-1})-\per(A_{i_*,x_{i_*}+d-1}).\]
\end{lemma}

\begin{proof}
The proof follows by developing the permanent along row $i$.
\end{proof}

\begin{lemma}
\label{lemma:the first lemma}
Let $i$ be such that $2 \leq x_i\leq r$, then
\[\per(\{x_k\})\leq \max\mathset{\per_i^+(\{x_k\}),\per_i^-(\{x_k\})}.\]
\end{lemma}

\begin{proof}
Assume the contrary, i.e., $\per(\{x_k\})> \per_i^+(\{x_k\})$  and  $\per(\{x_k\})> \per_i^-(\{x_k\})$. From Lemma \ref{stam} we get the following inequalities:
\begin{align*}
\per(A_{i^*,x_{i^*}+d})&<\per(A_{i^*,x_i^*})\\
%\label{eq:1}
\per(A_{i_*,x_{i_*}-1})&<\per(A_{i_*,x_{i_*}+d-1}).
%\label{eq:2}
\end{align*}
By Corollary \ref{cor:first cor} we get that
\begin{align*}
\per(A_{i^*,x_{i^*}})&\leq \per(A_{i^*,x_{i^*}-1})\\
%\label{eq:3}
\per(A_{i_*,x_{i_*}+d-1})&\leq \per(A_{i_*,x_{i_*}+d}).
%\label{eq:4}
\end{align*} 
Combining the four inequalities, it now follows that
\begin{align}
\per(A_{i_*,x_{i_*}-1})&<\per(A_{i_*,x_{i_*}+d-1})\leq \per(A_{i_*,x_{i_*}+d})\nonumber \\
&=\per(A_{i^*,x_{i^*}+d})\label{eq:6}\\
&<\per(A_{i^*,x_{i^*}})\leq \per(A_{i^*,x_{i^*}-1})\nonumber\\
&=\per(A_{i_*,x_{i_*}-1})\label{eq:9}
\end{align}
where equalities \eqref{eq:6} and \eqref{eq:9} follow from the fact
that $x_i=x_{i^*}=x_{i_*}$. Thus, we get a contradiction, and the
claim follows.
\end{proof}

\begin{lemma}
\label{lemma:push all the way}
When $i$ is such that $x_i<x_{i+1}\leq r $ and $\per(A)\leq \per_i^+(A)$, then
\[\per(A)\leq \per_i^+(A)\leq \per_i^{++}(A).\]
When $i$ is such that
$2\leq x_{i-1}< x_i$ and $\per(A)\leq \per_i^-(A)$, then
\[\per(A)\leq \per_i^-(A)\leq \per_i^{--}(A).\]
\end{lemma}

\begin{proof}
We start by proving the first claim. Define the $(0,1)$-matrix $B$
to be
\[B=(x_1,\dots,x_{i-1},x_i+1,x_{i+1},\dots,x_{d+r}),\]
and denote $j=(i+1)^*=\max\mathset{k:x_k=x_{i+1}}$. 

\paragraph{Case 1} Assume that
$x_{i+1}-x_i=1$, then by the definition of the operators
\[\per_i^+(\{x_k\})=\per(x_1,\dots,x_{i-1},x_i+1,x_{i+1},\dots,x_{d+r})=\per(B),\]
and also
\begin{align*}
\per_i^{++}(\{x_k\})&=\per(x_1,\dots,x_{i-1},x_i+1,x_{i+1},\dots,\\
&\qquad\qquad\qquad x_{(i+1)^*-1},x_{(i+1)^*}+1,x_{(i+1)^*+1},\dots,x_{d+r})\\
&=\per_j^+(B).
\end{align*}
By Lemma \ref{stam}, in order to prove the claim, i.e., $\per(B)\leq
\per_j^+(B)$, it suffices to show that $\per(B_{j,x_j})\leq
\per(B_{j,x_j+d})$. Since $\per(A)\leq \per_i^+(A)$, we conclude
\begin{equation}
\per(A_{i,x_i})\leq \per(A_{i,x_i+d}).
\label{eq:13}
\end{equation}
It is easy to see that
\begin{align*}
\per(A_{i,x_i})&=\per(B_{(i+1)^*,x_{(i+1)^*}-1})=\per(B_{j,x_j-1}),\\
\per(A_{i,x_i+d})&=\per(B_{(i+1)^*,x_{(i+1)^*}-1+d})=\per(B_{j,x_j-1+d}).
\end{align*}
Therefore, \eqref{eq:13} turns to
\begin{equation}
\per(B_{j,x_j-1})\leq \per(B_{j,x_j-1+d}).
\label{eq:14}
\end{equation}
By Corollary \ref{cor:first cor}
\begin{align}
\per(B_{j,x_j})&\leq \per(B_{j,x_j-1})
\label{eq:15}\\
\per(B_{j,x_j-1+d})&\leq \per(B_{j,x_j+d}).
\label{eq:16}
\end{align} 
Combining inequalities \eqref{eq:14}, \eqref{eq:15}, and \eqref{eq:16}, we get
\[\per(B_{j,x_j})\leq \per(B_{j,x_j-1})\leq \per(B_{j,x_j-1+d})\leq \per(B_{j,x_j+d}),\]
which proves the claim.

\paragraph{Case 2} Assume that $x_{i+1}-x_i\geq 2$. The proof in this
case is nearly identical. By definition,
\begin{align*}
\per_i^+(A)&=\per(x_1,\dots,x_{i-1},x_i+1,x_{i+1},\dots,x_{d+r})\\
\per_i^{++}(A)&=\per(x_1,\dots,x_{i-1},x_i+2,x_{i+1},\dots,x_{d+r}).
\end{align*}
Therefore, by Lemma \ref{stam}, in order to prove the claim it
suffices to show that
\[\per(A_{i,x_i+1})\leq \per(A_{i,x_i+d+1}).\]
From the fact that $\per(A)\leq \per_i^+(A)$, we conclude that
\begin{equation}
\per(A_{i,x_i})\leq \per(A_{i,x_i+d}).
\label{eq:18}
\end{equation} 
From Corollary \ref{cor:first cor}
\begin{align}
\per(A_{i,x_i+1})&\leq \per(A_{i,x_i}) \label{eq:19}\\
\per(A_{i,x_i+d})&\leq \per(A_{i,x_i+d+1}).
\label{eq:20}
\end{align}
Combining inequalities \eqref{eq:18}, \eqref{eq:19}, and \eqref{eq:20}, we get
\[\per(A_{i,x_i+1})\leq \per(A_{i,x_i})\leq \per(A_{i,x_i+d})\leq \per(A_{i,x_i+d+1}),\] 
and that completes the proof for the first claim.

The second claim of the lemma, again, easily follows from the symmetry
of the problem by rotating $A$.
\end{proof}

We are now in a position to prove the two main claims of the section. We
first calculate the value of the maximum permanent.
\begin{theorem}
\label{Th:bound on the perm}
Let $A\in M_{d+r}^d$, $0<r \leq d$, then
\[\per(A)=\binom{d-r}{\floorenv{\frac{d-r}{2}}}
\parenv{\floorenv{\frac{d+r}{2}}}!
\parenv{\ceilenv{\frac{d+r}{2}}}!.\]
Furthermore, the matrix $A=(\{x_i\})$ that satisfies
\begin{equation}
\label{eq:conf}
x_i=\begin{cases}
1 & 1\leq i \leq  \floorenv{\frac{d+r}{2}}, \\
r+1 & \text{otherwise}
\end{cases}
\end{equation}
is a member of $M^d_{d+r}$.
\end{theorem}

\begin{proof}
Let $A$ be a matrix that achieves the maximum permanent. If $A$ is not of
the required form there is at least one row,
$i$, that is not flushed to the right or left, i.e., $1<x_i<r+1$. By
Lemma \ref{lemma:the first lemma} we know that either 
$\per(A)\leq\per_i^+(A)$ or $\per(A)\leq \per_i^-(A)$.

Assume that $\per(A)\leq \per_i^+(A)$ (the proof for the other case is
symmetric). Since $\per(A)$ is the maximum achievable permanent,
necessarily $\per(A)=\per_i^+(A)$. It now follows by Lemma
\ref{lemma:push all the way} that
$\per(A)\leq \per_i^{++}(A)$.

By repeatedly using Lemma
\ref{lemma:push all the way} on the last block of $1$'s that was moved we
can continue to push blocks one step at a time, all in the same
direction. This procedure is terminated when we can no longer push
the block, i.e., when we have reached one of the matrix's edges. Thus,
we have reduced by one the number of blocks that are not flushed to
the right or left edges. We can therefore push all the blocks that are
not flushed to the edges until reaching some edge.

We conclude that the maximum permanent is also attained when
all the blocks are flushed to the edges. Let
$\lfloor \frac{d+r}{2}\rfloor + x$ be the number of blocks that are flushed to
the left edge, and thus, $\lceil \frac{d+r}{2}\rceil - x$ are flushed to the
right. We note that if $\lceil \frac{d+r}{2}\rceil-x < r$ then the permanent
is $0$, and so we can safely assume $\lceil \frac{d+r}{2}\rceil-x \geq r$.
Let us also denote by $\oodd$ the indicator function for
$d+r$ being odd. Thus, the permanent of this configuration is
\begin{align*}
&\parenv{\floorenv{\frac{d+r}{2}}+x}!
\parenv{\ceilenv{\frac{d+r}{2}}-x}!
\binom{d-r}{\floorenv{\frac{d+r}{2}}+x-r}=\\
&\qquad =(d-r)!\prod_{k=1}^r \parenv{\floorenv{\frac{d-r}{2}}+x+k}
\parenv{\ceilenv{\frac{d-r}{2}}-x+k}\\
&\qquad =(d-r)!\prod_{k=1}^r\parenv{\frac{d-r-\oodd}{2}+x+k}
\parenv{\frac{d-r+\oodd}{2}-x+k}\\
&\qquad =(d-r)!\prod_{k=1}^r \sparenv{\parenv{\frac{d-r}{2}+k}^2-
\parenv{x-\frac{\oodd}{2}}^2}.
\end{align*}
Hence, for $n=d+r$ even, the maximum is achieved when $x=0$, and for
$n=d+r$ odd, the maximum is achieved when $x=0,1$. In either case, 
when all the blocks are flushed to the edges of the matrix, the
maximum is achieved only when $\lfloor \frac{d+r}{2}\rfloor$ of the
blocks are flushed to one edge, and all the rest are flushed to the
other edge, and this completes the proof.
\end{proof}
Having proved the upper bound we want to know which matrix configurations achieve the bound with equality.

\begin{theorem}
\label{Th: classification}
Let $A\in M_{d+r}^d$, $0\leq r \leq d$, then the only possible configurations
of $A=(x_1,x_2,\dots,x_{d+r})$, 
up to a permutation of the rows and columns, are
\begin{equation}
\label{eq:optconf}
x_i = \begin{cases}
  1 & 1\leq i \leq \floorenv{\frac{d+r}{2}}\\
  r+1 & \ceilenv{\frac{d+r}{2}} < i \leq d+r.
\end{cases}
\end{equation}
Note that for $n=d+r$ odd, the value of $x_{\ceilenv{\frac{d+r}{2}}}$
is unconstrained.
\end{theorem}

\begin{proof}
For the case of $n=d+r$ even, assume to the contrary that there exists
$A\in M_{d+r}^d$ with a different configuration
than the claimed. By Theorem \ref{Th:bound on the perm}, we know that
we can push the non-zero blocks of $A$ along the rows without reducing the
permanent to achieve a matrix $A'$ with configuration
as in \eqref{eq:conf}. Let us denote the matrix before the last block
push as $A''$.
W.l.o.g., the configuration of
$A''=(\{x''_i\})$ is given by
\[x''_i = \begin{cases}
1 & 1\leq i < \frac{d+r}{2}\\
2 & i=\frac{d+r}{2}\\
r+1 & \text{otherwise.}
\end{cases}
\]
By our assumption, $\per(A)=\per(A')=\per(A'')$. However,
\begin{align*}
\per(A'')-\per(A') & =
\per(A''_{\frac{d+r}{2},d+1})-\per(A''_{\frac{d+r}{2},1})\\
& =\parenv{\frac{d+r}{2}-1}!\parenv{\frac{d+r}{2}}!
\sparenv{\binom{d-r}{\frac{d-r}{2}+1}-\binom{d-r}{\frac{d-r}{2}}}\\
& < 0,
\end{align*}
a contradiction. For the case of $n=d+r$ odd, the proof follows
the same logical steps but is more tedious as it has to consider more
cases, and is therefore given in \ref{ap:a}.
\end{proof}

\section{Asymptotic Results for $d<\frac{n}{2}$}
\label{sec:assymptotic results}
We now turn to show some asymptotic results for the case of
$d<\frac{n}{2}$. We follow the notation of Wanless \cite{Wan99}. With
$A\in \Gamma_n^d$, $A=(a_{i,j})$, we associate a bipartite graph
$G(A)$ with two vertex sets, $V=\mathset{v_1,v_2,\dots,v_n}$, which
represents the rows of $A$, and $U=\mathset{u_1,u_2,\dots,u_n}$, which
represents the columns of $A$. There is an edge $(v_i,u_j)$ iff
$a_{i,j}=1$. For every vertex $w\in V\cup U$ we denote by $N(w)$ its
set of neighbors, and its degree by $D(w)=\abs{N(w)}$. Finally, we
denote by $\oplus$ the direct-sum operator, and moreover, as in
\cite{Wan99}, we use $rA$ as shorthand for $A\oplus A \oplus\dots\oplus
A$ (where there are $r$ copies of $A$).

We are interested only in the
structure of matrices in $M_n^d$ up to isomorphism because the
permanent function is preserved under permutations of rows and
columns. We say that $\{C_i\}_{i=1}^k$ are the components of $A$ if $A
\cong C_1 \oplus C_2 \oplus\dots C_k $ and each $C_i$ is fully
indecomposable. Denote the order of a component, $C_i$, or a matrix,
$A$, by $\ord(C_i)$, and $\ord(A)$ respectively.

Our main results in this section are based heavily on the results of
\cite{Wan99}. We first mention a technical result from \cite{Wan99}
using the same notation. Define the following functions:
\begin{align*}
F(a,b)&=(a!)^\frac{b}{a},\\
D(k)&=\frac{F(k,1)}{F(k-1,1)},\\
C(k)&=\frac{D(k)}{D(k-1)},\\
B(k,v)&=C(k)^v((k - v)^2+2v(k - v)D(k - 1) + v(v - 1)(D(k - 1))^2).
\end{align*}
\begin{lemma}[{\cite[Lemma 1]{Wan99}}]
\label{wanless lemma}
For every integer $k\geq 3$ there exists
$\epsilon_k>0$ such that $B(k,v)<k^2- \epsilon_k$ for each integer $v$
satisfying $0<v<k$.
\end{lemma}

We will use another technical lemma:
\begin{lemma}[{\cite[p.~50]{AloSpe00}}]
\label{factorial inequality}
For every two integers $a,b$ satisfying $b\geq a+2>3$, the following
inequality holds
\[(a!)^{\frac{1}{a}}(b!)^{\frac{1}{b}}<((a+1)!)^{\frac{1}{(a+1)}}((b-1)!)^{\frac{1}{(b-1)}}.\]
\end{lemma}

We now turn to our specific setting and prove the following lemma.

\begin{lemma}
\label{stam lemma}
Let $B\in \Gamma_n^d$ be such that it does not contain $1_{d\times d}$
as a sub-matrix.  Let $W$ be a set of $2d$ contiguous column vertices,
i.e., $W=\mathset{u_i,u_{i+1},\dots,u_{i+2d-1}}\subseteq U$ for some
$i\in[n]$, then there is either some vertex $u_j\in W$ such that
$D(u_j)\neq d$ or there are two row vertices $v_x,v_y\in V$ such that
\begin{enumerate}
	\item $0<\abs{N(v_x)\cap N(v_y)}<D(v_x)$
	\item $D(u_k)=d$ for all $u_k\in N(v_x)\cup N(v_y)$. 
	\item $N(v_x),N(v_y)\subseteq W$.
\end{enumerate}
\end{lemma}
\begin{proof}
If there is some column vertex $u_j\in W$ such that $D(u_j)\neq d$ then
we are done. Otherwise, $D(u_j)=d$ for each $u_j\in W$.  Now, we know the
column vertex $u_{i+d}$ has degree $d$, and, by our assumption throughout the
paper that the identity permutation is in the anticode,
$v_{i+d}\in N(u_{i+d})$. On the other hand $B$ does not contain
$1_{d\times d}$ as a sub-matrix, and so there is a row vertex $v_j\in
N(u_{i+d})$ such that $N(v_{i+d})\cap N(v_j)\neq N(v_{i+d})$. Note
that $u_{i+d}\in N(v_{i+d})\cap N(v_j)$, and since the neighbors of
$v_{i+d}$ and $v_j$ form a contiguous block of column vertices, we get
that $$N(v_{i+d}),N(v_j)\subseteq W.$$ Now set $v_x=v_{i+d}$, $v_y=v_j$,
and the proof is complete.
\end{proof}
\begin{corollary}
\label{factorial cor}
For any integers $d$, $T$,and $n$,
where $T$ is even and $T\leq n$, the maximum of the function
$\prod_{i=1}^n F(x_i,1)$ subject to the constraints
\begin{enumerate}
	\item $\sum_{i=1}^{n}x_i=nd$.
	\item $x_i\geq 1$ are integers.
	\item \label{constraint 3} $T\leq \abs{\mathset{x_i:x_i\neq d}}$.
\end{enumerate}
is obtained exactly when the variables $x_i$ are as equal as possible, i.e., 
\[\abs{\mathset{x_i:x_i=d}}=n-T, \qquad 
\abs{\mathset{x_i:x_i=d+1}}=\abs{\mathset{x_i:x_i=d-1}}=\frac{T}{2}.\]
Therefore,
\[\prod_{i=1}^n F(x_i,1)\leq F(d,n-T)F(d-1,\frac{T}{2})F(d+1,\frac{T}{2}).\]
\end{corollary}

\begin{proof}
Recall that $F(x,1)=(x!)^{\frac{1}{x}}$. 
If there are two indices $i$ and $j$ such that $x_i\geq x_j +2$, then
by Lemma \ref{factorial inequality}, the value of
$\prod_{i=1}^n F(x_i,1)$ would increase if we add $1$ to $x_j$ and subtract 
$1$ from $x_i$, as long as we do not violate constraint \ref{constraint 3}.
\end{proof}

\begin{theorem}
\label{th:good theorem}
For each $A\in \Gamma_a^d$ there exists $m(A)\in\N$ such that
$\per(A\oplus t1_{d\times d} )>\per(B)$ for every integer $t$ such that 
$a + td>m(A)$ and every $B\in \Gamma_{a+td}^d$ which does not contain
$1_{d\times d}$ as a sub-matrix.
\end{theorem}

\begin{proof}
The claim is empty for $d=1,2$, since for $d=1$ there is nothing to
prove, and for $d=2$ there is no such $B\in \Gamma_{a+td}^d$ which
does not contain $1_{d\times d}$. Set $n=a+td$, then by Lemma
\ref{stam lemma} we know that for every
$l\in [\lfloor\frac{n}{2d}\rfloor]$ there is either a column vertex
$u_{i_l}\in\{u_{2d(l-1)+1},\dots,u_{2ld}\}$ such that $D(u_{i_l})\neq d$,
or there is a pair of row vertices, $v_{x_l}$ and $v_{y_l}$, such that
\begin{enumerate}
\item $0<\abs{N(v_{x_l})\cap N(v_{y_l})}<D(v_{x_l})$.
\item $D(u_k)=d$ for all $u_k\in N(v_{x_l})\cup N(v_{y_l})$.
\item $N(v_{x_l})\cup N(v_{y_l})\subseteq \{u_{2d(l-1)+1},\dots,u_{2ld}\}$.
\end{enumerate}

Let $M$ be the set of all $l\in [\lfloor \frac{n}{2d}\rfloor]$ such
that there exists a pair of row vertices, $v_{x_l}$ and $v_{y_l}$, as
above.  Set
\[T=\begin{cases}
\floorenv{\frac{n}{2d}}-|M| & \text{$\floorenv{\frac{n}{2d}}-|M|$ is even, } \\
\floorenv{\frac{n}{2d}}-|M|-1 & \text{otherwise.}
\end{cases}
\]
It is easy to see that $T+|M|\geq \lfloor \frac{n}{2d}\rfloor-1$. Note
that for any $l,k\in M$, $l\neq k$, 
\[(N(v_{x_l})\cup
N(v_{y_l}))\cap(N(v_{x_k})\cup N(v_{y_k}))=\emptyset.\]
For each pair
$(v_{x_l},v_{y_l})$, $l\in M$, let $a_l=\abs{N(v_{x_l})\cap N(v_{y_l})}$.  We
bound the permanent of $B$ from above by using the following steps:
\begin{enumerate}
\item 
  Expand $\per(B)$ along the row vertices $\mathset{v_{x_l},v_{y_l}}_{l\in M}$.
\item 
  Upper bound the expansion of the columns
  $\cup_{l\in M}(N(v_{x_l})\cup N(v_{y_l}))$ by using 
  Eq.~\eqref{bregman inequality}.
\item 
  Upper bound the expansion of the rest of the columns by using
  Eq.~\eqref{bregman inequality}, Corollary \ref{factorial cor}, and the
  fact that in these columns there are exactly
  $d(n-|\cup_{l\in M}(N(v_{x_l})\cup N(v_{y_l}))|)$
  non-zero entries, with at least $T$
  columns vertices with degree not equal to $d$.
\end{enumerate}
Therefore, for steps 1 and 2, the upper bound is
\begin{align}
\label{eq:23} 
&\prod_{l\in M}\Big[(d - a_l)^2 F(d - 1, 2d - 2a_l - 2)F(d - 2, a_l) \nonumber \\
&\qquad +2a_l (d - a_l)F(d - 1, 2d - 2a_l - 1)F(d - 2,a_l - 1)\nonumber \\
&\qquad +a_l (a_l - 1)F(d - 1, 2d - 2a_l)F(d - 2,a_l - 2)\Big].
\end{align}
the upper bound for step 3 using Corollary \ref{factorial cor} is
\begin{align}
\label{eq:26}
\nonumber &
F\parenv{d, n -T-\sum_{l\in M}(2d - a_l)}\parenv{(d+1)!^{\frac{1}{d+1}}(d-1)!^{\frac{1}{d-1}}}^{\frac{T}{2}}=\\
\nonumber &\qquad =
F\parenv{d, n-\sum_{l\in M}(2d - a_l)}\frac{\parenv{(d+1)!^{\frac{1}{d+1}}(d-1)!^{\frac{1}{d-1}}}^{\frac{T}{2}}}{d!^\frac{T}{d}}\\
&\qquad <
F\parenv{d, n-\sum_{l\in M}(2d - a_l)}\cdot\delta^\frac{T}{2},
\end{align}
where the last inequality follows from Lemma \ref{factorial inequality}
for some $0<\delta<1$. Combining \eqref{eq:23} and \eqref{eq:26} we get
\begin{align*}
\per(B) & \leq F(d,n)\delta^\frac{T}{2}\prod_{l\in M}
\sparenv{\frac{F(d-1,2d-2)}{F(d,2d)}\frac{F(d,a_l)F(d-2,a_l)}{F(d-1,2a_l)}}\cdot\\
& \quad\cdot\sparenv{(d - a_l)^2+2a_l (d - a_l)\frac{F(d - 1, 1)}{F(d - 2, 1)}+ a_l(a_l - 1)\frac{F(d - 1, 2)}{F(d - 2, 2)}},
\end{align*}
which in our notation becomes
\[\per(B)\leq F(d,n)\delta^\frac{T}{2}\prod_{l\in M}\frac{1}{d^2}B(d,a_l).\]
We know that $T+|M|\geq \lfloor \frac{n}{2d} \rfloor-1$, thus, by
Lemma \ref{wanless lemma} and by taking $t$, and hence $n$, large enough,
we can make $\per(B)$ be less than an arbitrary small fraction of
$F(d,n)$.

On the other hand, for $n=a+td$
\[\per(A\oplus t1_{d\times d})=(d!)^t\per(A)=F(d,n)\frac{\per(A)}{F(d,a)},\]
which is a constant fraction of $F(d,n)$ for any $t$. Hence, there
exists $m(A)$ such that if $a+td>m(A)$ then
$\per(B)<\per(A\oplus t1_{d\times d})$ as required.
\end{proof}

Though the set of matrices under study is different from the one
studied by Wanless \cite{Wan99}, the claim regarding their permanent
in Theorem \ref{th:good theorem} is exactly the same as the claim in
Theorem 1 in \cite{Wan99}. Thus, Theorems 3, 5, and 7 in \cite{Wan99},
which rely almost entirely on that claim, follow in our setting as well
with very slight adjustments. We
bring them here for completeness. For adjusted proofs, 
the reader is referred to \ref{ap:b}.

\begin{theorem}
\label{th:wan3}
For each integer $d$ there exists $b_d$ such that for any $n$ and any
$A\in M_n^d$, the largest component in $A$ that does not contain
$1_{d\times d}$ as a sub-matrix, is of order at most $b_d$.
\end{theorem}

Let $b^*_d$ denote the smallest integer with the property of
$b_d$ from Theorem \ref{th:wan3}.

\begin{theorem}
\label{th:wan5}
Let $d\leq n$ be positive integers. Every $A\in M_n^d$ is of the form
\[A\cong a1_{d\times d} \oplus C_1 \oplus C_2 \dots\oplus C_h\]
where $a\geq 0$ and $0\leq h \leq d-1$. Moreover $G(C_i)$ is connected,
$C_i \in M_{\ord (C_i)}^d$, and if in addition $C_i$ does not contain
$1_{d\times d}$ as a sub-matrix, then $\ord(C_i)\leq b^*_d$.
\end{theorem}

\begin{theorem}[{\cite[Theorem 7]{Wan99}}]
\label{th:wan7}
For each positive integer $d$ there exists $\mu_d$ such that $M_n^d$
is periodic for $n\geq \mu_d$ in the sense that $A\in M_n^d$ if and
only if $A\oplus 1_{d\times d} \in M_{n+d}^d$.
\end{theorem}

\section{Summary and Conclusions}
\label{sec:conclusion}
Motivated by new applications of error-correcting codes over
permutations under the $\ell_\infty$-norm, we have studied anticodes of
maximum size for the infinity metric. The results, together with the set-antiset
method, enable us to derive an improved upper bound on the size of
optimal codes (see \cite{TamSch10}). For $d > \frac{n}{2}$ we
classified all the optimal anticodes with maximal distance $d-1$, and
showed that their size is
\[\binom{2d-n}{\floorenv{\frac{2d-n}{2}}}
\parenv{\floorenv{\frac{n}{2}}}!\parenv{\ceilenv{\frac{n}{2}}}!.\]

For $d< \frac{n}{2}$, based on the results of \cite{Wan99}, we gave
asymptotic results on the structure of optimal anticodes. We showed
that for sufficiently large $n$, all but at most $d-1$ components of
any optimal anticode are $1_{d \times d}$. Moreover, some periodic
property of the optimal anticodes was shown.

It is tempting to combine all the results to the following conjecture:
\begin{conjec}
Denote $r=n \bmod d$, then for any $n$, the structure of the optimal
anticode of maximal distance $d-1$ is the set of permutations
$M=\{\sigma\in S_n:a_{i,\sigma(i)}=1,1\leq i \leq n\}$, where
\[
pA=(a_{i,j})=\parenv{\begin{array}{c@{}c@{}c@{}c@{}c}
\begin{array}{|c|}
\hline
1_{d\times d}\\
\hline
\end{array} & & & & \\
& \begin{array}{|c|}
\hline
1_{d\times d}\\
\hline
\end{array} & & \text{\Large 0} & \\
& & \ddots & & \\
& \text{\Large 0} & & \begin{array}{|c|}
\hline
1_{d\times d}\\
\hline
\end{array} & \\
& & & & 
\begin{array}{|c|}
\hline
P\\
\hline
\end{array}
\end{array}}.
\]
where along the diagonal we have $\floorenv{\frac{n}{d}}-1$ blocks of
$1_{d\times d}$, and $P$ is of the form given in \eqref{eq:optconf}.
It can then be easily seen that
\[\abs{M}=\per(A)=
(d!)^{\floorenv{\frac{n}{d}}-1}
\binom{d-r}{\floorenv{\frac{d-r}{2}}}
\parenv{\floorenv{\frac{d+r}{2}}}!\parenv{\ceilenv{\frac{d+r}{2}}}!.\]
\end{conjec}

\appendix
\section{Proof of Theorem \ref{Th: classification} -- Continued}
\label{ap:a}
We give here the proof of Theorem \ref{Th: classification} for the 
case of $n=d+r$ odd.

\begin{proof}
Let us consider the case of $n=d+r$ odd. First, we note that by
developing the permanent along the middle row, all matrices of
configuration \eqref{eq:optconf} have the same permanent regardless of
the value of $x_{\lceil \frac{d+r}{2} \rceil}$, i.e., the starting
column of the non-zero block in the middle row. Since one of these
configurations coincides with \eqref{eq:conf}, all matrices of
configuration \eqref{eq:optconf} have maximum permanent.

Assume to the contrary that there exists
$A\in M_{d+r}^d$ with a different configuration
than the claimed. By Theorem \ref{Th:bound on the perm}, we know that
we can push the non-zero blocks of $A$ along the rows without reducing the
permanent to achieve a matrix $A'$ with configuration
as in \eqref{eq:optconf}. Let us denote the matrix before the last block
push as $A''$.
W.l.o.g., the configuration of
$A''=(\{x''_i\})$ is given by
\[x''_i = \begin{cases}
1 & 1\leq i < \floorenv{\frac{d+r}{2}}\\
2 & i=\floorenv{\frac{d+r}{2}}\\
r+1 &  \lceil\frac{d+r}{2}\rceil < i \leq d+r.
\end{cases}\]
Note again that the value of $x_{\lceil\frac{d+r}{2}\rceil}$ is
unconstrained. 

By repeatedly using Theorem \ref{Th:bound on the perm} we can push the
non-zero block of row $\lceil\frac{d+r}{2}\rceil$ while maintaining the
maximum permanent value, until reaching a matrix $A^*=(\{x^*_i\})$ of
one of the two following configurations:
\[x^*_i = \begin{cases}
1 & 1\leq i < \floorenv{\frac{d+r}{2}}\\
2 & i=\floorenv{\frac{d+r}{2}}\\
2\text{ or } r+1 & i=\ceilenv{\frac{d+r}{2}}\\
r+1 &  \ceilenv{\frac{d+r}{2}} <i \leq d+r.
\end{cases}\]
Finally, let $(A^{**})=(\{x^{**}_i\})$ be defined by
\[x^{**}_i=\begin{cases}
1 & i= \floorenv{\frac{d+r}{2}} \\
x^{**}_i   & \text{otherwise.}
\end{cases}\]
We note that $A^{**}$ is of configuration \eqref{eq:optconf}, and
thus, of maximum permanent.  Therefore, by our assumptions,
\[\per(A)=\per(A')=\per(A'')=\per(A^*)=\per(A^{**}).\]
      
\paragraph{Case 1} Let $x^*_{\ceilenv{\frac{d+r}{2}}}=2$. One can readily
verify that
\begin{align*}
\per(A^*_{\floorenv{\frac{d+r}{2}},1})&=\parenv{\frac{d+r+1}{2}}!
\parenv{\frac{d+r-3}{2}}!\binom{d-r}{\floorenv{\frac{d-r}{2}}},\\
\per(A^*_{\floorenv{\frac{d+r}{2}},d+1})&=\parenv{\frac{d+r+1}{2}}!
\parenv{\frac{d+r-3}{2}}!\binom{d-r}{\floorenv{\frac{d-r}{2}}}\frac{d+r-3}{d+r+1}.
\end{align*}
It follows that
\[\per(A^*)-\per(A^{**})=
\per(A^*_{\floorenv{\frac{d+r}{2}},d+1})-
\per(A^*_{\floorenv{\frac{d+r}{2}},1}) < 0,\]
a contradiction.

\paragraph{Case 2} Let $x_{\ceilenv{\frac{d+r}{2}}}=r+1$.
Again, it is easily verifiable that
\begin{align*}
\per(A^*_{\floorenv{\frac{d+r}{2}},1})&=\parenv{\frac{d+r+1}{2}}!
\parenv{\frac{d+r-3}{2}}!\binom{d-r}{\floorenv{\frac{d-r}{2}}},\\
\per(A^*_{\floorenv{\frac{d+r}{2}},d+1})&=\parenv{\frac{d+r+1}{2}}!
\parenv{\frac{d+r-3}{2}}!\binom{d-r}{\floorenv{\frac{d-r}{2}}-1}.
\end{align*}
Once again, it follows that
\[\per(A^*)-\per(A^{**})=
\per(A^*_{\floorenv{\frac{d+r}{2}},d+1})-
\per(A^*_{\floorenv{\frac{d+r}{2}},1}) < 0,\]
a contradiction.
\end{proof}

\section{Proofs of Theorems \ref{th:wan3}, \ref{th:wan5}, and \ref{th:wan7}}
\label{ap:b}

The following are very slight adjustments to the proofs given by Wanless
in \cite{Wan99}. They are brought here for completeness.

\begin{proof}[Proof of Theorem \ref{th:wan3}]
For every integer $d\leq i < 2d$, choose some $A_i\in\Gamma_n^d$. Define
\[b_d=\max\mathset{m(A_d),m(A_{d+1}),\dots,m(A_{2d-1})}\]
where $m(A_i)$ was defined in Theorem \ref{th:good theorem}.

Assume to the contrary that $A\in M_n^d$ contains a component $C$
bigger than $b_d$ that does not contain $1_{d\times d}$ as a
sub-matrix. By Theorem \ref{th:good theorem} we can increase $\per(A)$
by replacing $C$ with $A_i\oplus t1_{d\times d}$, where
$d\leq i < 2d$, $i\equiv \ord(C) \bmod d$, and $t=(\ord(C)-i)/d$.
This contradiction proves the claim.
\end{proof}

\begin{proof}[Proof of Theorem \ref{th:wan5}]
Assume $A$ has $d$ connected components $C_1,\dots,C_d$. By Theorem
\ref{th:wan3}, the order of any component of $A$ not having $1_{d\times d}$
as a sub-matrix, is upper bounded by $b^*_d$. Let us now look at
the partial sums $s_i=\sum_{j=1}^i \ord(C_i)$. Obviously, either there is some
$j$ such that $s_j\equiv 0\pmod{d}$, or there are distinct $i$ and $j$
for which $s_i\equiv s_j\pmod{d}$. In any case, there are surely integers
$1\leq a\leq b\leq d$ for which $\sum_{i=a}^b\ord(C_i)=ld$ for some positive
integer $l$.

The permanent is multiplicative components, and therefore,
$C_i\in M_{\ord(C_i)}^d$. Furthermore, by Br{\'e}gman's Theorem,
$C_a\oplus \dots \oplus C_b\cong l1_{d\times d}$. Thus, $A$ has at most $d-1$
connected components which are not isomorphic to $1_{d\times d}$.
\end{proof}

\begin{proof}[Proof of Theorem \ref{th:wan7}]
For the first direction, assume $A\oplus 1_{d\times d}\in M_{n+d}^d$, then
every component maximizes its permanent and so $A\in M_n^d$. For the other
direction, assume $A\in M_n^d$ and $B\in M_{n+d}^d$. Further, let us assume
$n>(d-1)b^*_d$. We now have one of two cases:

\paragraph{Case 1} Either $B$ contains a connected component which
is $1_{d\times d}$, or all the connected components of $B$
do not contain $1_{d\times d}$ as a sub-matrix. Thus,
by using Theorems \ref{th:wan3} and \ref{th:wan5} in the latter case,
we are assured that
$B\cong 1_{d\times d}\oplus B'$ for some $B'\in\Gamma_n^d$. It now follows that
$\per(B')\leq \per(A)$ and so 
$\per(B)=\per(B'\oplus 1_{d\times d}\leq \per(A\oplus 1_{d\times d})$ and then
necessarily $\per(A\oplus 1_{d\times d})=\per(B)$, i.e., 
$A\oplus 1_{d\times d}\in M_{n+d}^d$.

\paragraph{Case 2} There exists a connected component $C$ of $B$ which contains
$1_{d\times d}$ as a sub-matrix. When viewed as a matrix of configuration
$C=(x_1,\dots,x_{\ord(C)})$, $x_{i+1}\geq x_i$, let us examine the top left
occurrence of $1_{d\times d}$ as a sub-matrix in $C$. By changing all the $1$'s
above and below the sub-matrix $1_{d\times d}$ to $0$'s, we get a matrix 
$C'\in\Gamma_{\ord(C)}^{\leq d}$, where $\Gamma_{n}^{\leq d}$ stands for the
set of $(0,1)$-matrices with exactly one contiguous non-zero block in each
row of size at most $d$. It is readily verifiable that $\per(C')=\per(C)$.

After the change, the matrix $B$ becomes $B'\in\Gamma_{n+d}^{\leq d}$ for which
$\per(B)=\per(B')$. In addition, 
$B'=1_{d\times d}\oplus B''$, where $B''\in\Gamma_{n}^{\leq d}$. We can, now,
arbitrarily change $0$'s to $1$'s in $B''$ so as to get a matrix
$B^*\in\Gamma_{n}^d$. Obviously, $\per(B'')\leq\per(B^*)\leq\per(A)$, and so
\[\per(B)=\per(B')=\per(B''\oplus 1_{d\times d})\leq\per(B^*\oplus 1_{d\times d})\leq\per(A\oplus 1_{d\times d}).\]
Just like in the previous case, it now follows that
$A\oplus 1_{d\times d}\in M_{n+d}^d$.
\end{proof}

\bibliographystyle{elsarticle-num}
%\bibliography{allbib}

\end{document}